\documentclass[manyauthors,nocleardouble,COMPASS]{cernphprep}
\usepackage{bm}
\usepackage{color}

\usepackage{amssymb}
\usepackage{bm}
\usepackage{amsmath}
\usepackage{amsbsy}
\usepackage{times}
\usepackage{epsfig} 
\usepackage{colordvi}
\usepackage{graphicx}
\usepackage{cite}
\usepackage{hyperref}
\usepackage{multicol}
\usepackage{booktabs}
\usepackage{ulem}
\usepackage{lineno}
\usepackage{lscape}
\usepackage{xcolor}

\RequirePackage[T1]{fontenc}

\newcommand{\rk}{R_{\rm K}}
\newcommand{\rp}{R_{\rm p}}

\begin{document}


\begin{titlepage}

\PHnumber{version 0.9}
\PHdate{\today}  
\title{Antiproton over proton and K$^-$ over K$^+$ \\ multiplicity ratios at high $z$ in DIS}

\date{}
\Collaboration{The COMPASS Collaboration}
\ShortAuthor{The COMPASS Collaboration}

\begin{abstract} \label{abstract}

The $\bar{\rm p} $ over p multiplicity ratio 
is measured in deep-inelastic scattering for the first time using (anti-) protons carrying a large fraction of the virtual-photon energy, $z>0.5$.
The data were obtained by the COMPASS
Collaboration using a 160 GeV muon beam impinging on an isoscalar $^6$LiD target. The regime of deep-inelastic
scattering is ensured by requiring $Q^2$ > 1 (GeV/$c$)$^2$ for the photon virtuality and 
$W > 5$ GeV/$c^2$ for the
invariant mass of the produced hadronic system. 
The range in Bjorken-$x$ is restricted to 
$0.01 < x < 0.40$.
Protons and antiprotons are identified in the momentum range 
$20 \div 60$ GeV/$c$. 
In the whole studied $z$-region, 
the $\bar{\rm p}$ over p multiplicity ratio
is found to be below  the lower limit expected from calculations based on leading-order perturbative Quantum Chromodynamics (pQCD).
Extending our earlier analysis of the K$^-$ over K$^+$ multiplicity ratio by including now events with larger virtual-photon energies,
this ratio becomes closer to the expectation of next-to-leading order pQCD.
The results of both analyses
strengthen our earlier conclusion that
the phase space available for hadronisation should be
taken into account in the pQCD formalism.

\end{abstract}

\vfill
\Submitted{(to be submitted to PLB)}

\end{titlepage}

{\pagestyle{empty}  
%
%
\section*{The COMPASS Collaboration}
\label{app:collab}
\renewcommand\labelenumi{\textsuperscript{\theenumi}~}
\renewcommand\theenumi{\arabic{enumi}}
\begin{flushleft}
M.G.~Alexeev\Irefnn{turin_u}{turin_i},
G.D.~Alexeev\Irefn{dubna}, 
A.~Amoroso\Irefnn{turin_u}{turin_i},
V.~Andrieux\Irefnn{cern}{illinois},
V.~Anosov\Irefn{dubna}, 
A.~Antoshkin\Irefn{dubna},
K.~Augsten\Irefnn{dubna}{praguectu}, 
W.~Augustyniak\Irefn{warsaw},
C.D.R.~Azevedo\Irefn{aveiro},
B.~Bade{\l}ek\Irefn{warsawu},
F.~Balestra\Irefnn{turin_u}{turin_i},
M.~Ball\Irefn{bonniskp},
J.~Barth\Irefn{bonniskp},
R.~Beck\Irefn{bonniskp},
Y.~Bedfer\Irefn{saclay},
J.~Berenguer~Antequera\Irefnn{turin_u}{turin_i},
J.~Bernhard\Irefnn{mainz}{cern},
M.~Bodlak\Irefn{praguecu},
F.~Bradamante\Irefn{triest_i},
A.~Bressan\Irefnn{triest_u}{triest_i},
M.~B\"uchele\Irefn{freiburg},
V.~E.~Burtsev\Irefn{tomsk},
W.-C.~Chang\Irefn{taipei},
C.~Chatterjee\Irefnn{triest_u}{triest_i},
M.~Chiosso\Irefnn{turin_u}{turin_i},
A.~G.~Chumakov\Irefn{tomsk},
S.-U.~Chung\Irefn{munichtu}\Aref{B}\Aref{B1}
A.~Cicuttin\Irefn{triest_i}\Aref{C}
P.~M.~M.~Correia\Irefn{aveiro},
M.L.~Crespo\Irefn{triest_i}\Aref{C}
% ...
\end{flushleft}
%
%
\begin{Authlist}
\item \Idef{aveiro}{University of Aveiro, Dept.\ of Physics, 3810-193 Aveiro, Portugal}
\item \Idef{bochum}{Universit\"at Bochum, Institut f\"ur Experimentalphysik, 44780 Bochum, Germany\Arefs{Q}$^,$\Arefs{R}}
\item \Idef{bonniskp}{Universit\"at Bonn, Helmholtz-Institut f\"ur  Strahlen- und Kernphysik, 53115 Bonn, Germany\Arefs{Q}}
\item \Idef{bonnpi}{Universit\"at Bonn, Physikalisches Institut, 53115 Bonn, Germany\Arefs{Q}}
\item \Idef{brno}{Institute of Scientific Instruments of the CAS, 61264 Brno, Czech Republic\Arefs{S}}
\item \Idef{calcutta}{Matrivani Institute of Experimental Research \& Education, Calcutta-700 030, India\Arefs{T}}
\item \Idef{dubna}{Joint Institute for Nuclear Research, 141980 Dubna, Moscow region, Russia\Arefs{T1}}
\item \Idef{freiburg}{Universit\"at Freiburg, Physikalisches Institut, 79104 Freiburg, Germany\Arefs{Q}$^,$\Arefs{R}}
\item \Idef{cern}{CERN, 1211 Geneva 23, Switzerland}
\item \Idef{liberec}{Technical University in Liberec, 46117 Liberec, Czech Republic\Arefs{S}}
\item \Idef{lisbon}{LIP, 1649-003 Lisbon, Portugal\Arefs{U}}
\item \Idef{mainz}{Universit\"at Mainz, Institut f\"ur Kernphysik, 55099 Mainz, Germany\Arefs{Q}}
\item \Idef{miyazaki}{University of Miyazaki, Miyazaki 889-2192, Japan\Arefs{V}}
\item \Idef{moscowlpi}{Lebedev Physical Institute, 119991 Moscow, Russia}
\item \Idef{munichtu}{Technische Universit\"at M\"unchen, Physik Dept., 85748 Garching, Germany\Arefs{Q}$^,$\Arefs{D}}
\item \Idef{nagoya}{Nagoya University, 464 Nagoya, Japan\Arefs{V}}
\item \Idef{praguecu}{Charles University, Faculty of Mathematics and Physics, 12116 Prague, Czech Republic\Arefs{S}}
\item \Idef{praguectu}{Czech Technical University in Prague, 16636 Prague, Czech Republic\Arefs{S}}
\item \Idef{protvino}{State Scientific Center Institute for High Energy Physics of National Research Center `Kurchatov Institute', 142281 Protvino, Russia}
\item \Idef{saclay}{IRFU, CEA, Universit\'e Paris-Saclay, 91191 Gif-sur-Yvette, France\Arefs{R}}
\item \Idef{taipei}{Academia Sinica, Institute of Physics, Taipei 11529, Taiwan\Arefs{W}}
\item \Idef{telaviv}{Tel Aviv University, School of Physics and Astronomy, 69978 Tel Aviv, Israel\Arefs{X}}
\item \Idef{triest_u}{University of Trieste, Dept.\ of Physics, 34127 Trieste, Italy}
\item \Idef{triest_i}{Trieste Section of INFN, 34127 Trieste, Italy}
\item \Idef{turin_u}{University of Turin, Dept.\ of Physics, 10125 Turin, Italy}
\item \Idef{turin_i}{Torino Section of INFN, 10125 Turin, Italy}
\item \Idef{tomsk}{Tomsk Polytechnic University, 634050 Tomsk, Russia\Arefs{Y}}
\item \Idef{illinois}{University of Illinois at Urbana-Champaign, Dept.\ of Physics, Urbana, IL 61801-3080, USA\Arefs{Z}}
\item \Idef{warsaw}{National Centre for Nuclear Research, 02-093 Warsaw, Poland\Arefs{a} }
\item \Idef{warsawu}{University of Warsaw, Faculty of Physics, 02-093 Warsaw, Poland\Arefs{a} }
\item \Idef{warsawtu}{Warsaw University of Technology, Institute of Radioelectronics, 00-665 Warsaw, Poland\Arefs{a} }
\item \Idef{yamagata}{Yamagata University, Yamagata 992-8510, Japan\Arefs{V} }
\end{Authlist}
%
%
\renewcommand\theenumi{\alph{enumi}}
\begin{Authlist}
\item [{\makebox[2mm][l]{\textsuperscript{\#}}}] Corresponding authors
\item \Adef{B}{Also at Dept.\ of Physics, Pusan National University, Busan 609-735, Republic of Korea}
\item \Adef{B1}{Also at Physics Dept., Brookhaven National Laboratory, Upton, NY 11973, USA}
\item \Adef{C}{Also at Abdus Salam ICTP, 34151 Trieste, Italy}
\item \Adef{D}{Supported by the DFG cluster of excellence `Origin and Structure of the Universe' (www.universe-cluster.de) (Germany)}
\item \Adef{F}{Supported by the DFG Research Training Group Programmes 1102 and 2044 (Germany)}
\item \Adef{G}{Also at Chubu University, Kasugai, Aichi 487-8501, Japan}
\item \Adef{H}{Also at Dept.\ of Physics, National Central University, 300 Jhongda Road, Jhongli 32001, Taiwan}
\item \Adef{I}{Also at KEK, 1-1 Oho, Tsukuba, Ibaraki 305-0801, Japan}
\item \Adef{J}{Present address: Universit\"at Bonn, Physikalisches Institut, 53115 Bonn, Germany}
\item \Adef{K}{Also at Moscow Institute of Physics and Technology, Moscow Region, 141700, Russia}
\item \Adef{L}{Also at Yerevan Physics Institute, Alikhanian Br. Street, Yerevan, Armenia, 0036}
\item \Adef{M}{Also at Dept.\ of Physics, National Kaohsiung Normal University, Kaohsiung County 824, Taiwan}
\item \Adef{M1}{Supported by ANR, France with P2IO LabEx (ANR-10-LBX-0038) in the framework ``Investissements d'Avenir'' (ANR-11-IDEX-003-01)}
\item \Adef{N}{Also at Institut f\"ur Theoretische Physik, Universit\"at T\"ubingen, 72076 T\"ubingen, Germany}
\item \Adef{N1}{Retired}
\item \Adef{N2}{Present address: Brookhaven National Laboratory, Brookhaven, USA}
\item \Adef{O}{Also at University of Eastern Piedmont, 15100 Alessandria, Italy}
\item \Adef{O1}{Present address: Universit\"at Hamburg, 20146 Hamburg, Germany}
\item \Adef{P}{Present address: RWTH Aachen University, III.\ Physikalisches Institut, 52056 Aachen, Germany}
\item \Adef{Q}{Supported by BMBF - Bundesministerium f\"ur Bildung und Forschung (Germany)}
\item \Adef{R}{Supported by FP7, HadronPhysics3, Grant 283286 (European Union)}
\item \Adef{S}{Supported by MEYS, Grant LM20150581 (Czech Republic)}
\item \Adef{T}{Supported by B.~Sen fund (India)}
\item \Adef{T1}{Supported by CERN-RFBR Grant 12-02-91500}
\item \Adef{U}{Supported by FCT, Grants CERN/FIS-PAR/0007/2017 and  CERN/FIS-PAR/0022/2019 (Portugal)}
\item \Adef{V}{Supported by MEXT and JSPS, Grants 18002006, 20540299, 18540281 and 26247032, the Daiko and Yamada Foundations (Japan)}
\item \Adef{W}{Supported by the Ministry of Science and Technology (Taiwan)}
\item \Adef{X}{Supported by the Israel Academy of Sciences and Humanities (Israel)}
\item \Adef{Y}{Supported by the Russian Federation  program ``Nauka'' (Contract No. 0.1764.GZB.2017) (Russia)}
\item \Adef{Z}{Supported by the National Science Foundation, Grant no. PHY-1506416 (USA)}
\item \Adef{a}{Supported by NCN, Grant 2017/26/M/ST2/00498 (Poland)}
\end{Authlist}

\clearpage
}

\maketitle

\section{Introduction} \label{sec:int}
Within the standard approach of perturbative Quantum Chromodynamics
(pQCD), hadron production from an active quark 
in a deep-inelastic scattering process (DIS) is effectively described by non-pertur\-ba\-tive
objects called fragmentation functions (FFs).
These functions presently cannot be predicted by theory,
but their scale evolution is described by the DGLAP equations \cite{dglap}.
For a given negative four-momentum transfer squared $Q^2$,
in leading order (LO) pQCD the FF $D^{\rm h}_{\rm q}(z, Q^2)$ 
represents the  
probability density 
that a hadron h is produced
in the fragmentation of a quark with flavour q.
The produced hadron carries
a fraction $z$ of the virtual-photon energy $\nu$, 
where the latter is defined in the laboratory frame.

The cleanest way to access FFs consists in studying
single-inclusive hadron production in lepton
annihilation, ${\rm e}^+ + {\rm e}^- \to {\rm h}+$X,
where the remaining final state X is not analysed. However, only information about $D_{\rm
q}^{\rm h}+D_{\bar{\rm q}}^{\rm h}$
is accessible there and only limited flavour separation is possible. Additional input, like
semi-inclusive measurements of deep-inelastic lepton-nucleon scattering (SIDIS),
is required
to fully understand quark fragmentation into hadrons. In the case of the SIDIS cross
section, fragmentation functions are convoluted with parton distribution functions (PDFs).
As these are rather well known, fragmentation functions for q and $\bar{\rm q}$ can be
accessed separately and full flavour separation is possible in principle. As a result,
fragmentation functions obtained using only e$^+$e$^-$ data differ in some cases 
significantly from those that
were determined by additionally taking 
into account data from SIDIS or other processes, see
Refs.~\cite{hkns07,hkks16,jam01,dss01,dss02,nnpdf}.

Recently, the HERMES and COMPASS Collaborations have
published several papers concerning unidentified hadron, 
pion and kaon multiplicities in SIDIS, see Refs. \cite{comp_pi, comp_K, comp_hpt, hermes}. 
In the most recent COMPASS article~\cite{comp_Kratio} 
it was shown that for kaons at high $z$  
the K$^-$ over K$^+$ multiplicity ratio $\rk$ falls below the lower limit predicted by pQCD. 
From the measured $\nu$-dependence
it was concluded that in experiments with similar or lower 
centre-of-mass energy than in COMPASS
an insufficient description of the data by pQCD may affect 
the high-$z$ region.
This kinematic region is important in many respects, as
$e.g.$ transverse-momentum-dependent azimuthal asymmetries are 
quite pronounced there~\cite{tmds}.
Hence the above described phenomenon should be better understood
in order to avoid possible bias 
when extracting fragmentation functions
and/or transverse-momentum-dependent PDFs and FFs
by applying the naive pQCD formalism to SIDIS data in the high-$z$ region. 

In order to provide more experimental input for further phenomenological studies,
we present here for the first time the COMPASS
results on the $\bar{\rm p}$ over p 
multiplicity ratio $\rp$ at high $z$, $i.e.$ $z>0.5$, 
which are obtained from SIDIS data 
taken on an isoscalar target. In addition we 
present new results on 
$\rk$, obtained in a $\nu$-range extended with respect to Ref.~\cite{comp_Kratio}, which became attainable by improving
the kaon identification procedure.
Note that when measuring 
a multiplicity ratio, several systematic uncertainties
cancel in both theory and experiment. Thus a multiplicity ratio 
can be considered as 
one of the most robust observables presently available 
when analysing SIDIS data.

This Letter is organised as follows. In Section \ref{sec:th}, 
pQCD-based predictions for  $\rp$ and $\rk$
are discussed.
Experimental set-up and data selection
are described in Section \ref{sec:expdata}.
The analysis method is presented in Section \ref{sec:ana},
followed by the discussion of systematic uncertainties in
Section \ref{sec:sys}.
The results are presented and discussed in Section \ref{sec:res}.

\section{Theoretical framework and model expectations} \label{sec:th}

Hadrons of type h produced in the final state of DIS
are commonly characterised by their relative abundance. The hadron multiplicity
$M^{\rm h}$ is defined as ratio of the SIDIS cross section for
hadron type h and
the cross section for an inclusive measurement of the deep-inelastic scattering process (DIS):
\begin{equation}
\label{mult_def}
     \frac{{\rm d} M^{\rm h}(x,Q^2,z)}{{\rm d}z} =\frac{{\rm d}^3\sigma^{\rm h}(x,Q^2,z)/{\rm d}x {\rm d}
Q^2 {\rm d} z}{{\rm d}^2\sigma^{\rm DIS}(x,Q^2)/{\rm d}x {\rm d} Q^2}.
\end{equation}
Here, $x$ denotes the Bjorken scaling variable.
The cross sections $\sigma^{\rm DIS}$ and $\sigma^{\rm h}$
can be composed using the standard factorisation approach of pQCD \cite{nlo_form,dsv}.
In the following, the LO pQCD expressions for the cross section calculations will be used.
In the LO approximation for the multiplicity, where
the sum over parton species $a= {\rm q},  \bar{\rm q}$ is 
weighted by the square of the electric charge $e_a$ of the quark
expressed in units of the elementary charge, only simple products 
of PDFs $f_a(x,Q^2)$ and FFs $D_a^{\rm h}(z,Q^2)$ are involved instead
of the aforementioned convolutions:
\begin{equation}
\label{mult_LO}
    \frac{{\rm d} M^{\rm h}(x,Q^2,z)}{{\rm d}z}=\frac{\sum\limits_{a}{e_a^2f_a(x,Q^2)D_a^{\rm h}(z,Q^2)}}
    {\sum\limits_{a}{e_a^2f_a(x,Q^2)}}.
\end{equation}
For a deuteron target, the 
$\bar{\rm p}$ over p
multiplicity ratio in LO pQCD reads as follows:
\begin{equation} \label{eq:main0}
\rp (x,Q^2,z)= \frac{{\rm d}M^{\bar{\rm p}}\!(x,Q^2,z)/{\rm d}z}{{\rm d} M^{{\rm p}}\!(x,Q^2,z)/{\rm d}z}=
\frac{4.5(\bar{\rm u}+ \bar{\rm d}) D_{\rm fav} +({\rm 5u + 5d + 2{\rm s} + 2\bar{\rm s} }) D_{\rm unf} }
     {4.5( {\rm u + d}) D_{\rm fav} +  (5 \bar{\rm u} + 5 \bar{\rm d} + 2{\rm s} + 2 \bar{\rm s}) D_{\rm unf}}.
\end{equation}
Here, u, $\bar{\rm u}$, d, $\bar{\rm d}$, s, $\bar{\rm s}$
denote the PDFs in the proton for corresponding quark flavours.
Their dependences on $x$ and $Q^2$ are omitted for brevity.
The symbols $D_{\rm fav}$ ($D_{\rm unf}$) denote favoured (unfavoured) fragmentation functions and their dependence on $z$ and $Q^2$ are also omitted for brevity.
 Presently, proton FFs and their ratios 
are not well known at high $z$ as their extraction 
is based on e$^+$e$^-$ annihilation data only \cite{hkns07}.
Following Refs.~\cite{hkns07} and \cite{diquark} it is assumed that 
$D_{\rm u}^{\rm p}= 2D_{\rm d}^{\rm p} =D_{\rm fav}$.
In addition, the existing data do not allow to distinguish
between different functions $D_{\rm unf}$
for different quark flavours.
In the large-$z$ region, the ratios $D_{\rm unf}/D_{\rm fav}$ 
are expected to be small \footnote{For kaons, this expectation is indeed confirmed 
in pQCD fits already at moderate values of $z$, see $e.g.$ Ref. \cite{dss02}.}. 
Neglecting $D_{\rm unf}$ in Eq.~(\ref{eq:main0}) leads to
the following lower limit for $\rp$ in LO pQCD
\begin{equation} \label{eq:main1}
\rp >  \frac{\bar{\rm u}+ \bar{\rm d} }{{\rm u+ d}},
\end{equation}
which depends only upon rather well known PDFs, and is independent
on the assumption that $D_{\rm u}^{\rm p}= 2D_{\rm d}^{\rm p} =D_{\rm fav}$. 
It is interesting to notice that the value of the lower limit
predicted by LO pQCD is the same for both protons and kaons, 
see Ref.~\cite{comp_Kratio}. 
However, one expects $\rk>\rp$ as in the case of kaons 
the strange quark FFs $(D_{\rm s}^{\rm K^{-}}, D_{\bar{\rm s}}^{\rm K^{+}})$
are of the favoured type, contrary to the proton case. The expected value of
$\rk/\rp$ is about 1.10 when using 
FFs from Ref.~\cite{dss01} 
and the MSTW08LO PDF set from Ref.~\cite{pdf_mstw08}, 
where the strange-quark contribution is suppressed with
respect to the light-quark sea.
It can be as large as 1.15 if 
strange quarks are not suppressed with respect
to the light-quark sea as suggested by the interpretation of some LHC measurements~\cite{pdf_mmht14, pdf_nnpdf}. 
On the other hand, newer kaon FFs that are available only at NLO~\cite{dss02}
suggest that the ratio of $D_{\rm \bar{s}}^{K^{+}}/D_{\rm u}^{K^{+}}$ is about 1.5 times smaller than
originally obtained in Ref.~\cite{dss01}. 
In this case, $\rk/\rp$ is reduced back to about 1.10. 
Therefore, $\rk/\rp=1.10\pm 0.05$
appears as
a reasonable expectation based on the LO pQCD formalism.

The present analysis is performed in two $x$-bins, below and above $x=0.05$.
The average values of $x$ and $Q^2$ are 
$\langle x \rangle = 0.023$, $\langle Q^2 \rangle$ = 2.4 (GeV/$c)^2$ in the first $x$-bin and
$\langle x \rangle = 0.10$, $\langle Q^2 \rangle$ = 9.8 (GeV/$c)^2$ in the second one.
Based on Eq.~(\ref{eq:main1}) and the MSTW08LO PDF set,
the expected lower limits on $\rp$ 
in these two $x$-bins are 0.51 and 0.28.
These values are
about 10\% higher if newer PDF sets as in Refs.~\cite{pdf_mmht14, pdf_nnpdf} are used instead. 
Due to the above mentioned lack of reliable proton FFs at NLO, presently no predictions can be made for the
lower limit of $\rp$ at higher perturbative order.

We also evaluate
$\rp$ with the LEPTO Monte Carlo event generator \cite{lepto} (version 6.5),
with the result that the LUND string fragmentation model \cite{lund} used in LEPTO 
is incapable to model $\rp$ correctly.
For example, for $z\approx 0.5$ LEPTO predicts $\rp \approx 1$,
which is definitely not supported by the data as it will be shown below.
On the other hand, for $z>0.85$
the predicted value of $\rp$ falls below the naive LO pQCD lower limit. 
This is possible as in the LUND model 
the mechanism of string hadronisation
does not only depend on quark and hadron types and on $z$, as in the pQCD formalism,  
but also on the type of the target nucleon and on $x$,
see Ref. \cite{Aram} for more details.

Due to different lower momentum limits for particle identification at COMPASS, 
18 GeV/$c$ for protons and 9 GeV/$c$ for kaons, 
the observed $x$ and $Q^2$ distributions  are slightly different for pions and kaons. 
As a result, the lower limit on $R_{\rm K}$ is about 0.47,
which is obtained for $\langle x \rangle = 0.03$ and $\langle Q^2 \rangle$ = 1.6 (GeV/$c)^2$.
The LO pQCD predictions for the lower limit on $R_{\rm K}$ are $\nu$ independent,
because they depend on PDFs in the same way as given
in Eq.~(\ref{eq:main1}) for the proton case.
However, in our earlier measurement~\cite{comp_Kratio} a clear $\nu$ dependence was observed.
With higher values of $\nu$ accessible in the current measurement, we expect
the results to be in better agreement with the expectation of (N)LO pQCD.
We also note that the NLO lower limit for $\rk$ turns out
to be 10$\div$15\%
smaller than the LO pQCD lower limit given above, see Ref.
\cite{comp_Kratio}.

Some phenomenological models~\cite{TMC_old, TMC_ChL, TMC_new}
are able to accommodate $\rk$ below the pQCD limits presented 
above, but the predicted
effect is too small to explain our earlier published 
results~\cite{comp_Kratio}.
There are also important theoretical efforts ongoing to improve
the formalism 
(higher-order corrections, treatment of heavy quarks {\it etc.}),
see $e.g.$ Refs.~\cite{WVog_res,Ref2_1, Ref2_2, Ref2_3, Ref2_4, Ref2_5}, which however 
do not affect the interpretation of the data shown 
in Ref.~\cite{comp_Kratio} and in the present paper.

\section{Experimental set-up and data selection} \label{sec:expdata}

The present analysis is based on COMPASS data taken in 2006.
The 160 GeV/$c$ $\mu^+$ beam
delivered by the M2 beam line of the CERN SPS had a momentum spread of 
about 5\%.
The beam was naturally polarised, but the polarisation
is not affecting this analysis since 
we integrate over 
azimuthal angle and transverse momentum of the produced hadrons.
The $^6$LiD target has a total length of 120 cm, which corresponds to  
about half of a hadron interaction length.
It is considered to be isoscalar, and 
the 0.2\% excess of neutrons over protons due to
the presence of additional material in the target ($^3$He and $^7$Li) is neglected.
The target was longitudinally polarised, but in the present analysis
the data are averaged over the target polarisation,
which leads to a remaining average
target polarisation below 1\%.

The COMPASS two-stage spectrometer has a polar-angle acceptance of $\pm$180 mrad,
and it is capable of detecting charged particles with momenta as low as 0.5 GeV/$c$.
However, in this analysis typical particle momenta are above 20 GeV/$c$.
The ring-imaging Cherenkov detector (RICH) was used to identify pions, kaons and protons.
Its radiator volume was filled with C$_4$F$_{10}$
leading to a threshold for pion, kaon and proton identification of about 3 GeV/$c$, 9 GeV/$c$ and
18 GeV/$c$ respectively.
Two trigger types are used in the analysis.
The ``inclusive'' trigger is based on a signal from a combination of hodoscope signals
caused by the scattered muon. The ``semi-inclusive'' trigger requires an energy deposition in one of
the hadron calorimeters.
The experimental set-up is described in more detail in Ref.~\cite{nim}.

The data selection criteria are kept similar
to those used in the recently published analyses~\cite{comp_K, comp_Kratio} whenever possible.
In order to formally ensure the applicability of the pQCD formalism,
the DIS region is selected  
by requiring $Q^2 > 1$ (GeV/$c)^2$ and $W > 5$ GeV/$c^2$
for the invariant mass of the produced hadronic system.
The fraction of the incoming muon energy carried by the virtual photon, $y$, is kept
larger than 0.1 to avoid the region
with degraded momentum resolution.

For the proton multiplicity analysis, 
the constraint $x>0.01$ is used in order to make the kinematic
coverage more similar to that of our earlier kaon studies 
~\cite{comp_Kratio}.       
In the present analysis, we study protons carrying a large fraction $z$ of the virtual-photon energy,
$z>0.5$. 
In order to ensure efficient proton identification by the RICH, only events
with proton momentum above 20 GeV/$c$ are used, $i.e.$ 2 GeV/$c$ above the RICH proton threshold. 
The upper limit for proton identification is set to 60 GeV/$c$.
Purity and efficiency of the proton selection are optimised  by imposing appropriate constraints 
on the likelihoods of
proton, kaon, pion and background hypotheses that are calculated 
by the RICH particle-identification software~\cite{rich_pid}. 

In our earlier studies of $\rk$ ~\cite{comp_Kratio},
kaons with
momenta between 12 GeV/$c$ and 40 GeV/$c$ were analysed for $z>0.75$.
By the improvements in the RICH particle-identification software described in Section \ref{sec:ana},
the momentum range extends now up to 55 GeV/$c$, 
which leads to a significant extension of the available $\nu$ range.
All other kaon selection criteria remain unchanged with respect to the earlier analysis.

\section{Analysis method} \label{sec:ana}

The proton (kaon) multiplicities \mbox{$M^{\rm p(K)}$($x$, $Q^2$, $z$)}
are determined from the
proton (kaon) yields $N^{\rm p(K)}$ normalised by the number of DIS events,
$N^{\rm DIS}$, and corrected by the acceptance $A^{\rm p(K)}(x,Q^2,z)$:
\begin{equation} \label{eq:ExpMul}
\frac{\text{d}M^{\rm p (K)}(x,Q^2,z)}{\text{d}z} =
\frac{1}{N^{\rm DIS}(x,Q^2)}\frac{\text{d}N^{\rm p (K)}(x,Q^2,z)}{\text{d}z} \frac{1}{A^{\rm p(K)}(x,Q^2,z)}\,.
\end{equation}

As in our earlier kaon analysis~\cite{comp_Kratio}, we use ``semi-inclusive'' triggers. 
This is possible because a
bias-free determination of $N^{\rm DIS}$ is not needed, as
the latter cancels in $\rp$ and $\rk$.
The total number of protons and anti-protons used in the analysis is about 50\,000.
In addition to about 64\,000 kaons analysed in Ref.~\cite{comp_Kratio}, 
there are about 13\,000 kaons more 
in the newly explored kinematic range. 
Note that the kinematic range for protons is wider than that for kaons. 

As it was mentioned in Section \ref{sec:th}, 
the proton analysis is performed in two $x$-bins, below and above $x=0.05$.
In each $x$-bin, nine bins
are used in the reconstructed $z$ variable $z_{\rm rec}$,
with the bin limits
0.50, 0.55, 0.60, 0.65, 0.70, 0.75, 0.80, 0.85, 0.90 and 1.10.
In addition, for events in the first
$x$-bin the data are separated
in four bins of proton momentum $p_{\rm h}$, with the
bin limits 20 GeV/$c$, 30 GeV/$c$, 40 GeV/$c$, 50 GeV/$c$ and 60 GeV/$c$. 
This 2-dimensional
binning allows implicit studies of the $\nu$-dependence of $\rp$. 
For the second $x$-bin, the
anti-proton statistics is too limited to perform the analysis in the additional dimension of (anti-)proton momentum. 
For kaons the present analysis is only performed for $x<0.05$,
using five $z$-bins with bin limits 0.75, 0.80, 0.85, 0.90, 0.95 and 1.05,
and three momentum bins with bin limits 40 GeV/$c$, 45 GeV/$c$, 50 GeV/$c$ and 55 GeV/$c$. 

In order to determine the multiplicity ratio $\rp$ from the raw yield of $\bar{\rm p}$ and p,
only a few correction factors have to be taken into account. First, the correction related to 
RICH efficiencies is 
applied. From an analysis of 
$\Lambda^0$ and $\overline{\Lambda^0}$ decays into an
(anti-)proton-pion pair it was concluded that the RICH efficiency for p is charge-symmetric within 
a precision of about 1\%. The  
proton selection, which was improved with respect to our earlier papers, 
ensures that the contamination from $\pi$ 
and K can be safely neglected. Upper limits to such a possible contamination are taken into account 
in the systematic uncertainty. The acceptance correction factors $A^{\rm p}$ for p and 
$\bar{\rm p}$ are determined using Monte Carlo simulations. The same unfolding method is 
used as in Ref.~\cite{comp_Kratio}, 
$i.e.$, in a given ($x$, $Q^2$) bin we 
calculate the ratio of the number of reconstructed events to that of generated ones. 
Note that in order to count generated (reconstructed) events, 
generated (reconstructed) variables are used. 
As for $z$ unfolding, we present the results as a function of $z_{corr}$,
which denotes the value of $z$ reconstructed in the experiment, corrected by the average 
difference between the generated and reconstructed values of $z_{rec}$, where the latter are determined by Monte Carlo simulations. The average acceptance ratio for the first $x$-bin is $A_{\bar{\rm p}}/A_{\rm p}= 0.912 \pm 0.004$ (stat.)
and a similar value is obtained for the second $x$-bin. 
The systematic uncertainty related to the acceptance ratio is 
discussed in the next section. 
It is also verified  by using the  DJANGOH Monte Carlo generator~\cite{djangoh} 
that in the COMPASS kinematics the radiative correction for positive and negative 
particles is of the same value within uncertainties, thus it cancels in the ratio.

Compared to the above proton analysis and the kaon analysis presented in Ref.~\cite{comp_Kratio},
the raw K$^ \pm$ yields are obtained in a different way, which is described below. 
After that,
the present analysis follows closely the same procedure as in the case
of the proton analysis and the one from Ref. \cite{comp_Kratio}.  
With respect to the proton analysis described above it is in addition verified
using simulations that 
the contamination from diffractive vector meson
decays ($e.g.$ $\phi \rightarrow {\rm K} ^+{\rm K}^-$)
and charm meson decays is negligible. 

\begin{figure}
\centerline{\includegraphics[clip,width=1.0\textwidth]{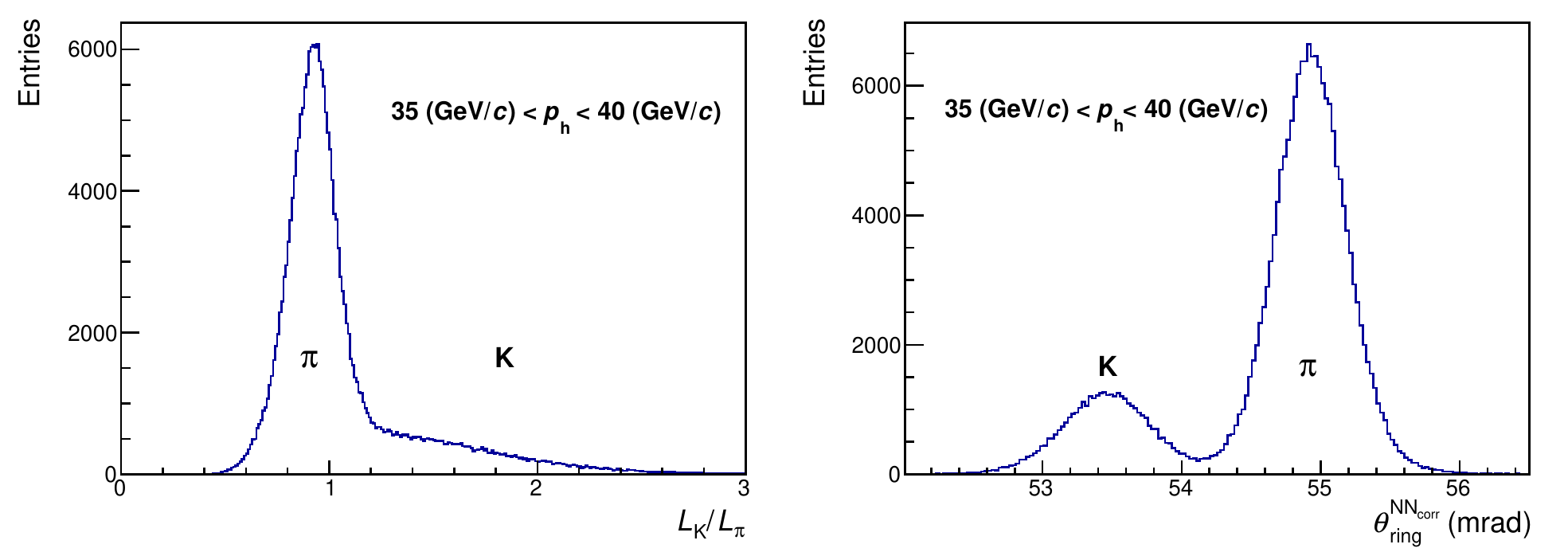}}
\caption{Left panel: RICH likelihood ratio of K over $\pi$ hypotheses for tracks
with momenta between 35 GeV/$c$ and 40 GeV/$c$ where the separation between K and $\pi$
is not obvious. In order to select 
kaons, the constraint 
$L_{\rm K}/L_{\pi}>1.5$ was used in Ref.~\cite{comp_Kratio}.
Right panel: reconstructed angle of Cherenkov photons in the ``ring fit'' after the 
NN correction (which is used in the present analysis), for the same data as shown in the left panel. Here, a much cleaner separation between
$\pi$ and K is visible.}
\label{fig:ana1}
\end{figure}

In the proton analysis and in the kaon analysis from Ref.~\cite{comp_Kratio}, 
the raw yields are obtained directly
by counting the number of events 
that fulfil certain criteria of RICH particle identification. 
However, by improving 
the RICH particle-identification software 
a better separation between $\pi$ and K
can be achieved at higher momenta.
For the present analysis, 
the polar angle $\theta$ of the Cherenkov photon rings is corrected by a Neural Networks (NN) parametrisation, which intends to improve the internal description of
the RICH sub-structure with respect to what was known during the original data production and the reconstruction.
This correction depends upon various track parameters like
position and angle at the RICH entrance, momentum of the particle {\it etc}. 
and is applied on an event-by-event basis. 
In the left panel of Fig.~\ref{fig:ana1},
we recall from our earlier analysis~\cite{comp_Kratio}
the likelihood ratio for the K/$\pi$ hypothesis
in the highest momentum bin, where the separation was
most challenging.
In order to optimise the uncertainties of $\rk$, a
lower limit of 1.5
had to be used there.
Using in the present analysis the NN method, the separation of kaons and pions is
improved considerably as illustrated
in the right panel of Fig.~\ref{fig:ana1}, where the 
$\theta$ distribution
after the NN correction is shown
for the same events as in the left panel.
A much better separation of the two particle species is clearly visible, which allows us
to extend the analysis to higher momenta up to 55 GeV.

In order to obtain the raw kaon yield, the spectra as the one shown
in the right panel of Fig.~\ref{fig:ana1} 
are fitted in each $z$ and $p_{\rm h}$ bin using  
the functional form described below.
It turns out that a single Gaussian
to describe the kaon peak and two Gaussians 
for the pion peak are sufficient to obtain the raw kaon yield.
The fit	is performed simultaneously in	all $z$ and $p_{\rm h}$ bins. 
This procedure is a source of non-negligible systematic uncertainties, 
especially
at higher $z$ and higher momenta. The systematic uncertainty related to this extraction
is described in Section \ref{sec:sys}. 

\section{Studies of systematic uncertainties} \label{sec:sys}

This section is split into two parts. In the first part, 
studies of systematic effects for 
the proton results are described. 
This is a rather standard analysis that benefits
from the significant knowledge acquired with the previously published COMPASS 
analyses~\cite{comp_pi, comp_K, comp_Kratio}. 
In the second part, the kaon results are described. 
As for the first time in COMPASS
a new method is used to estimate
the kaon yield, detailed studies are performed 
to verify the reliability of the results.
Additionally, 
standard studies as done for $\rk$ in Ref.~\cite{comp_Kratio} are also performed.

\subsection{Systematic uncertainties for $\rp$}

i) The COMPASS data taking was divided into periods,
mainly depending upon the schedule of the SPS accelerator. 
 A typical data period 
took about one week, and in between two periods
interventions to the COMPASS spectrometer could happen. 
The whole 2006 data taking 
took about half a year. 
Therefore, 
it is  verified that the values for $\rp$ obtained
from different data periods agree with one another.

ii) As in Ref.~\cite{comp_Kratio}, and contrary to standard multiplicity
analyses~\cite{comp_pi,comp_K,comp_hpt}, two trigger types are used
in this analysis, with or without
the requirement of energy deposit in the calorimeters. 
It is verified that
these two trigger types give 
consistent values for $\rp$. This
result is expected as 
for the lowest proton energy  analysed (20 GeV) calorimeter efficiencies are already close to 100\%.

iii) The key correction factor that has to be applied to the raw value of $\rp$ is the acceptance
difference between p and $\bar{\rm p}$. 
The COMPASS spectrometer is charge symmetric 
at the level of 1\%. 
However, protons and anti-protons interact differently with the target material as they 
do not have the same
re-interaction length in the long solid-state COMPASS $^6$LiD target.
Therefore, as already mentioned in Sect.~\ref{sec:ana}, the acceptance for $\bar{\rm p}$ 
is about 10\% lower than that  for p, with an estimated uncertainty of about 3\%.

iv) More complex methods of unfolding the acceptance were tested in 
Ref.~\cite{comp_Kratio}, as well as in the present analysis. 
They are giving very similar results when compared 
to the selected method, but their resulting covariance matrix has large off-diagonal elements. On the 
contrary, for the selected method the results in each bin and their statistical uncertainties can 
be considered to be 
independent from each other.

v) A correction factor 
has to be taken into account because of possibly different 
RICH reconstruction efficiencies
for p and $\bar{\rm p}$. 
While the correction factor is found
to be one, the systematic studies suggest that its uncertainty is
about 5\%. This uncertainty on $\rp$ is by
2\% larger than that found for $\rk$,
mostly due to the higher mass of the proton 
compared to that of the kaon, which leads 
to less photons per ring in the RICH in most
of the phase space region covered.
On top of that, some performed tests
are limited in precision
due to the small statistics, especially for anti-protons
at larger momenta and/or larger values of $z$.

vi) As in previous studies, the stability of $\rp$ is tested on data using several variables that are defined in the spectrometer coordinate system.
A clear instability is seen in the dependence 
of $\rp$ upon the azimuthal angle measured in the laboratory frame, as it was the case 
in our earlier analysis of $\rk$.
In Ref.~\cite{comp_Kratio}, this asymmetry
led to a systematic uncertainty of up to 12\%
in both $x$-bins.
In this analysis, for data binned in $x$ and $z$, 
the systematic uncertainty amounts
up to 5\% for the 1st $x$-bin and up to 11\% in the 2nd $x$-bin.
For data binned in $z$ and $p_{\rm h}$, it can be up to 15\% for high momenta. 
Thus in a significant part of the phase space
this systematic uncertainty is the dominant one.  

The total systematic uncertainty of $\rp$ is obtained by 
adding in quadrature the above discussed 
contributions. The relative systematic uncertainty is found
to range between 6\% and 16\%. The correlation between systematic uncertainties in 
various $z$ and $p_{\rm h}$-bins is about 0.7--0.8, as in Ref.~\cite{comp_Kratio}.

\subsection{Systematic uncertainties for $\rk$}
Most studies of systematic effects for kaon results follow
closely the ones from Ref. \cite{comp_Kratio}, which are also described
above for protons.
The systematic uncertainty related to the acceptance ratio
and the RICH efficiency ratio for the two kaon charges is taken
as in Ref.~\cite{comp_Kratio}, $i.e.$ 
2\% and 3\%, respectively.
The uncertainty related to the azimuthal-angle distribution of hadrons in the spectrometer  
is studied using the same method
as in our previous paper and the resulting relative uncertainty ranges between 4\% and 12\%. 
Compared to the analysis presented in Ref. \cite{comp_Kratio},
a new type of systematic uncertainty has to be studied,
which is related to the new method of extracting the 
raw kaon yields from RICH data.

First,
it is verified that the results obtained
with the new method do agree with those previously published~\cite{comp_Kratio}. Various combinations of functional forms are used
in the fit, $e.g.$ the main results are obtained using a Gaussian functional form to fit
the polar-angle distribution of the kaon and two Gaussian functions for the one of the pion.
In the systematic studies, we use a single or two Gaussian function(s)
for each particle type. 
With three Gaussian functions to describe the
polar-angle distribution of photons in the RICH detector,
there are nine free parameters in every single $z$ and hadron momentum bin, and for each of the two hadron charges. 
The fit in certain bins (at large $z$ and large momentum) results in very large
uncertainties on the obtained values of $\rk$. In order to improve accuracy, studies
are performed to determine which parameters can be kept common for the two charges and across various $z$ and momentum bins. 
For example, the pion and kaon Cherenkov opening angles depend only
on the particle momenta but not on $z$. 
Indeed, it is confirmed in the fit that this angle is independent
on $z$ within uncertainties. Altogether,
the initial 450 free parameters in the fit are reduced
by about a factor of three.
The systematic uncertainty of the final results on $\rk$
is evaluated by performing several fits, in which the number of free parameters is reduced by releasing certain constraints.

As a systematic uncertainty, half of the difference between maximum
and minimum value of $\rk$ obtained in these studies is taken.
The resulting 
relative uncertainty of the kaon yield
is found to range between 4\% and 25\%.
The total systematic uncertainty of $\rk$ is found to range between 7\% and 28\% of 
the $\rk$ value, and correspondingly between 0.4 and 1.1 of 
the statistical uncertainty on $\rk$.
As in previous analyses, the correlation
between systematic uncertainties in various
$z$ and $p_h$-bins is about 0.7--0.8.
We note that a fit of all data simultaneously may introduce correlations between $\rk$ values in different $z$ and $p_h$-bins.
These correlations are found to be below 5\% and hence neglected.

\section{Results and discussion} \label{sec:res}

\begin{figure}
\centerline{\includegraphics[clip,width=0.9\textwidth]{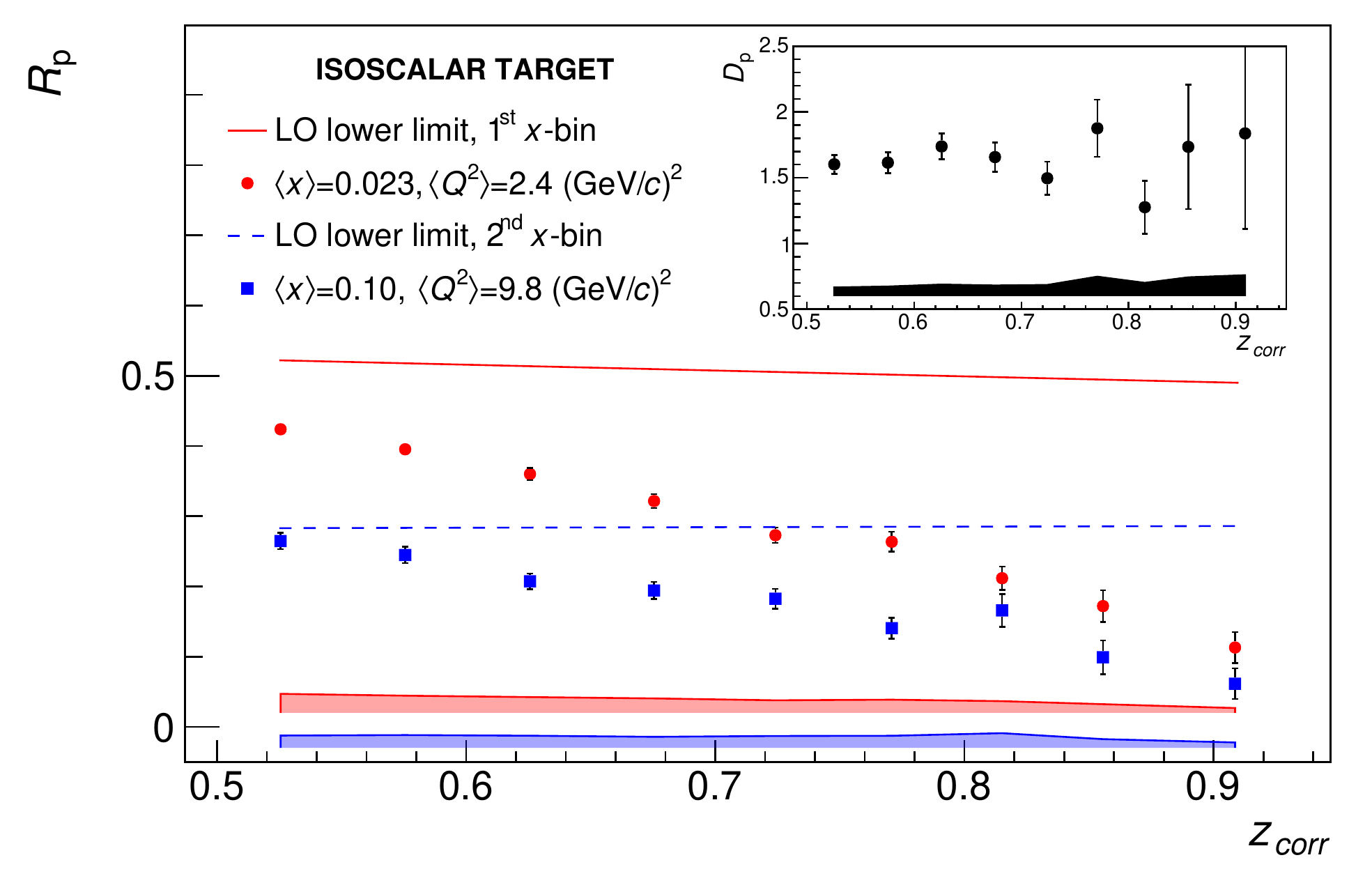}}
\caption{
Results on $\rp$ as a function
of $z_{\rm corr}$ for the two $x$-bins. The insert shows the double ratio $D_{\rm p}$
defined as the ratio of $\rp$ in the first $x$-bin over $\rp$ in the second $x$-bin.
Statistical uncertainties are shown by error bars and 
systematic uncertainties by 
shaded bands at the bottom.
The lines indicate the %
lower limit on $\rp$ predicted by LO pQCD 
using the PDF set from Ref.~\cite{pdf_mstw08}. The relative uncertainty of the limit
is below 4\% in both $x$-bins.
}
\label{fig:res1}
\end{figure}

In Fig.~\ref{fig:res1} and Table~\ref{tab:res0}, 
the results on the anti-proton over proton multiplicity ratio $\rp$ are presented
as a function of the variable $z_{\rm corr}$
for the two 
$x$-bins used in this analysis. 
The measured $z$-dependence of $\rp$ can be fitted
in both $x$-bins by simple functional forms, $e.g.$ $
\propto (1-z)^{\beta}$. The obtained $\beta$ value for this 
fit, $\beta=0.75\pm 0.04$, agrees 
within uncertainties well with $\beta=0.71\pm 0.03$ obtained from the fit 
to $\rk$ in Ref.~\cite{comp_Kratio}. Presently,
it is not clear if this  observed agreement between 
kaons and protons is 
accidental or not.
A ``double ratio'' $D_{\rm p}=R_{\rm p}(x<0.05)/R_{\rm p}(x>0.05)$,
is shown in the insert of the figure.
It may be considered  constant within 
uncertainties over
the full measured $z$-range, 
with an average value of $D_{\rm p}=1.62\pm 0.04_{\rm stat.}\pm 0.07_{\rm syst.}$.

\begin{figure}
\centerline{\includegraphics[clip,width=0.9\textwidth]{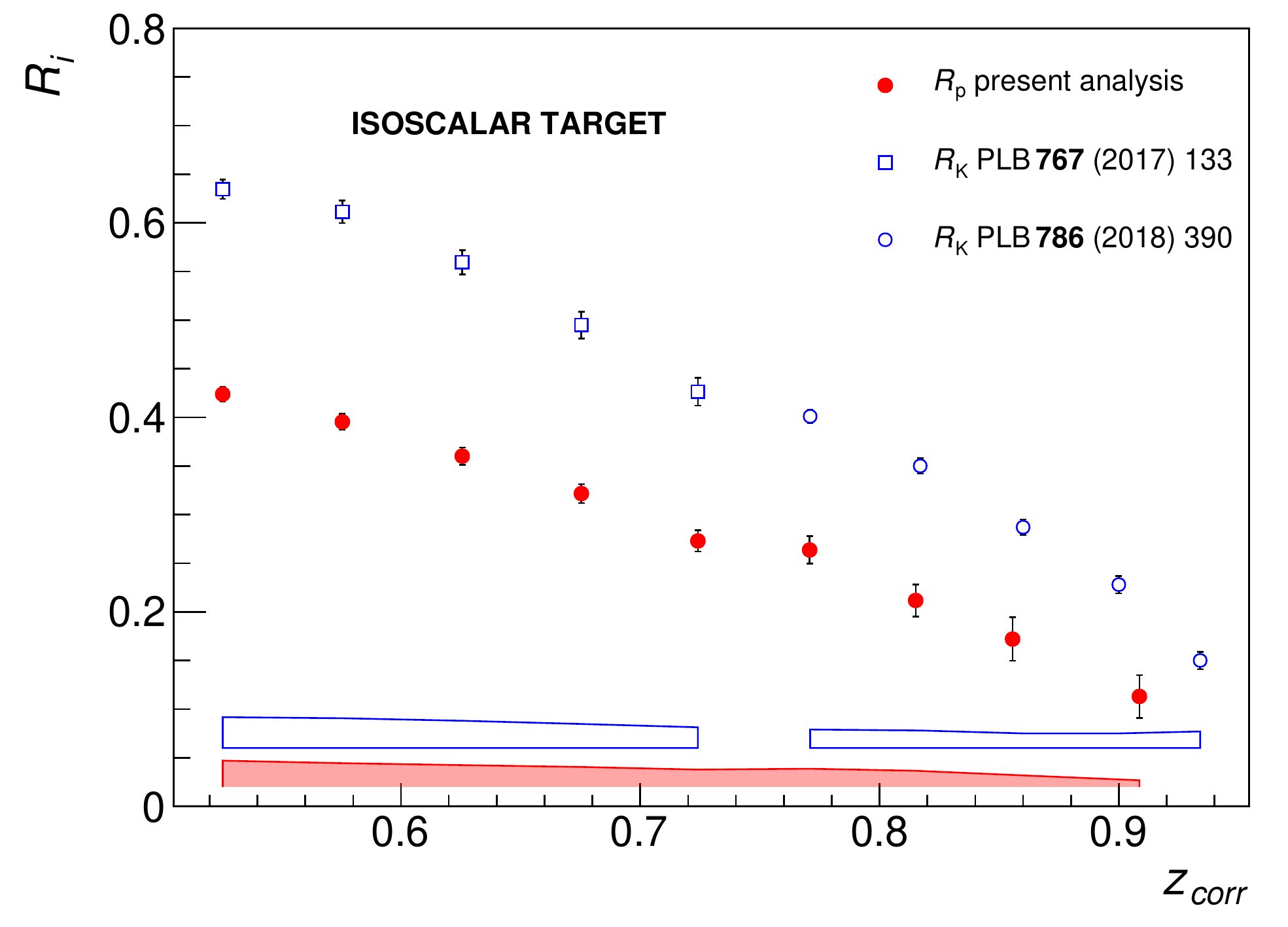}}
\caption{Results on $\rp$ and $\rk$ as a function
of $z_{\rm corr}$ for the first $x$-bin, $x<0.05$. The ratio 
$\rp$ falls 
below $\rk$ in the whole measured phase space. The kaon data come from Refs.~\cite{comp_K, comp_Kratio}.
Statistical uncertainties are shown by error bars, systematic uncertainties by the bands at the bottom.}
\label{fig:res2}
\end{figure}

The most important observation is that with the increase of $z$ the measured value of $\rp$ is 
increasingly undershooting the LO pQCD expectation,
which is 
0.51 and 0.28 calculated for the average kinematics of the data in the 1st and 2nd $x$-bin, respectively. 
It is remarkable that $\rp$ falls 
below the LO pQCD prediction over 
the whole measured $z$ range, which starts in this analysis from $z>0.5$. 
This effect was observed for $\rk$ only
for $z>0.8$. In Fig. \ref{fig:res2}, 
the comparison of 
$\rp$ with $\rk$ calculated using data in Ref.~\cite{comp_K} and from Ref. \cite{comp_Kratio}
shows that over the whole measured phase space 
$\rp$ falls 
significantly below $\rk$. 
As mentioned above,
the $x$ and $Q^2$ distributions 
from the two analyses are different, which can change the results by about 5-10\%.
We hence avoid to quote precise results on the $\rp/\rk$ ratio here. 
As discussed in Section~\ref{sec:th}, the lower limit for $\rp$ and $\rk$
in LO pQCD is the same.
For the ratio itself, a small difference of the order of 10\% is expected
due to the presence of favoured strange-quark fragmentation in the kaon case. 
The two effects, $i.e.$ different $x$ and $Q^2$ 
distributions  and favoured 
strange quark fragmentation in the case of kaons,
act in opposite directions. Thus in 
naive LO pQCD one would expect the proton and kaon data
points shown in Fig. \ref{fig:res2} to agree within better than 5\%, which is clearly not the case. 
This indicates that the additional correction to the pQCD formalism 
we suggested in Ref.~\cite{comp_Kratio},
which takes into account the phase space available for hadronisation, depends on the mass of the produced hadron.

\begin{figure}
\centerline{\includegraphics[clip,width=0.9\textwidth]{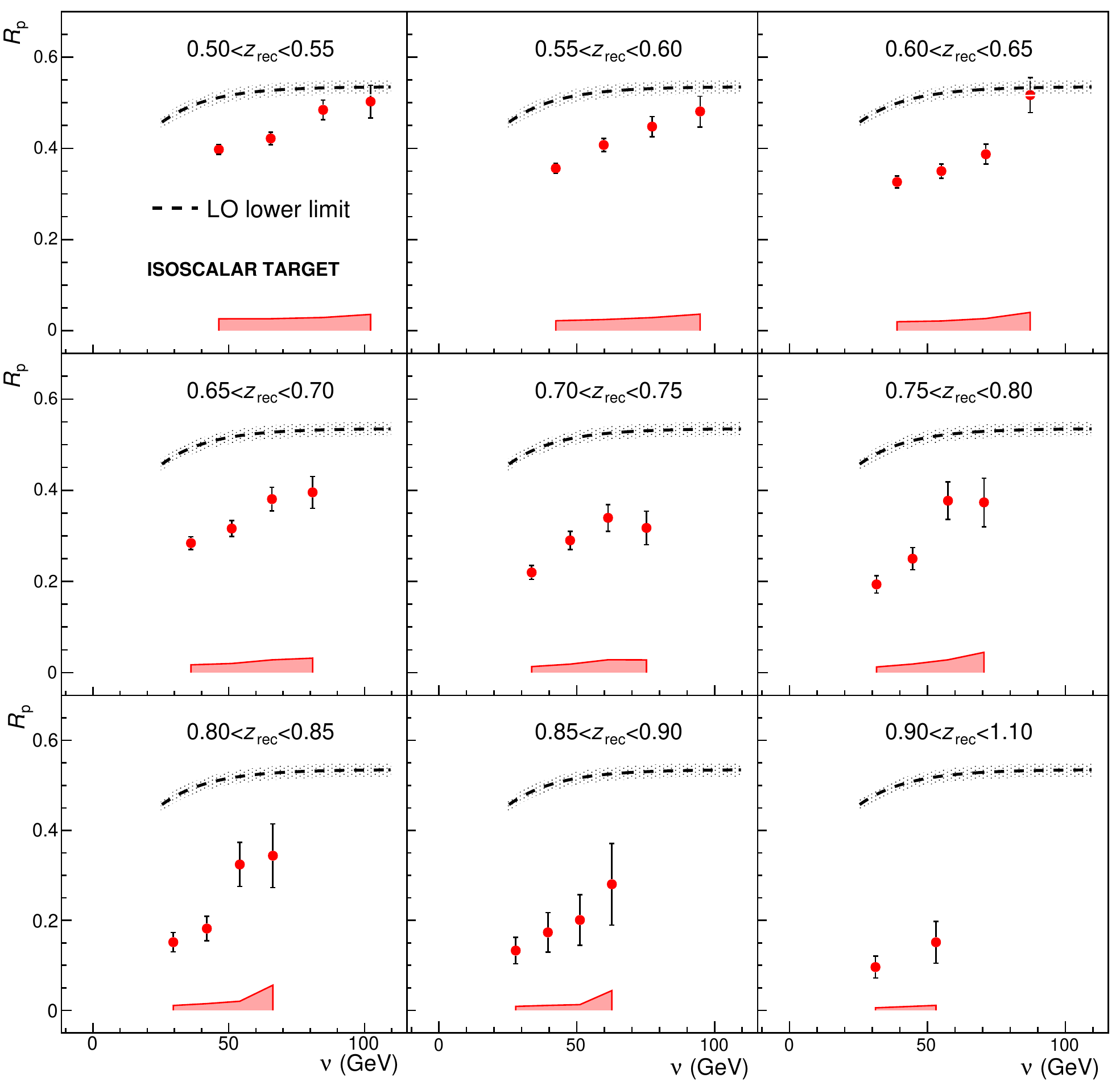}}
\caption{Results on $\rp$ as a function of $\nu$ in nine bins 
of $z_{\rm rec}$ for the first $x$-bin, $x<0.05$. Statistical uncertainties are shown by error bars, 
systematic uncertainties by the shaded bands at the bottom.
The curves represent the
lower limits for $\rp$ calculated in LO pQCD using \cite{pdf_mstw08} PDF set.
The shaded bands around the LO lower limits indicates
their uncertainty.
}
\label{fig:res2a}
\end{figure}

One of the striking features of the observed disagreement between the expectation
of (N)LO pQCD and the 
results on $\rk$ obtained  in Ref.~\cite{comp_Kratio} was the observed strong
dependence of $\rk$ on the virtual-photon energy $\nu$,
with values of $\rk$ closer to the pQCD prediction for higher $\nu$.
Our present results on $\rp$ 
do confirm a similar dependence for the proton case.
These results as well as the prediction of LO pQCD are shown
in Fig.~\ref{fig:res2a} and in Table~\ref{tab:res1}. Much higher energies
than those available in COMPASS
seem to be required to eventually reach in the high-$z$ region the lower limit of $\rp$ 
predicted by LO pQCD.
We mention that the lower limit of $\rp$ does not
directly depend on $\nu$. The $\nu$-dependence of the pQCD lower limit seen in Fig.~\ref{fig:res2a} 
is related to different mean values of $x$ and $Q^2$ for different values of $\nu$.

\begin{figure}
\centerline{\includegraphics[clip,width=0.9\textwidth]{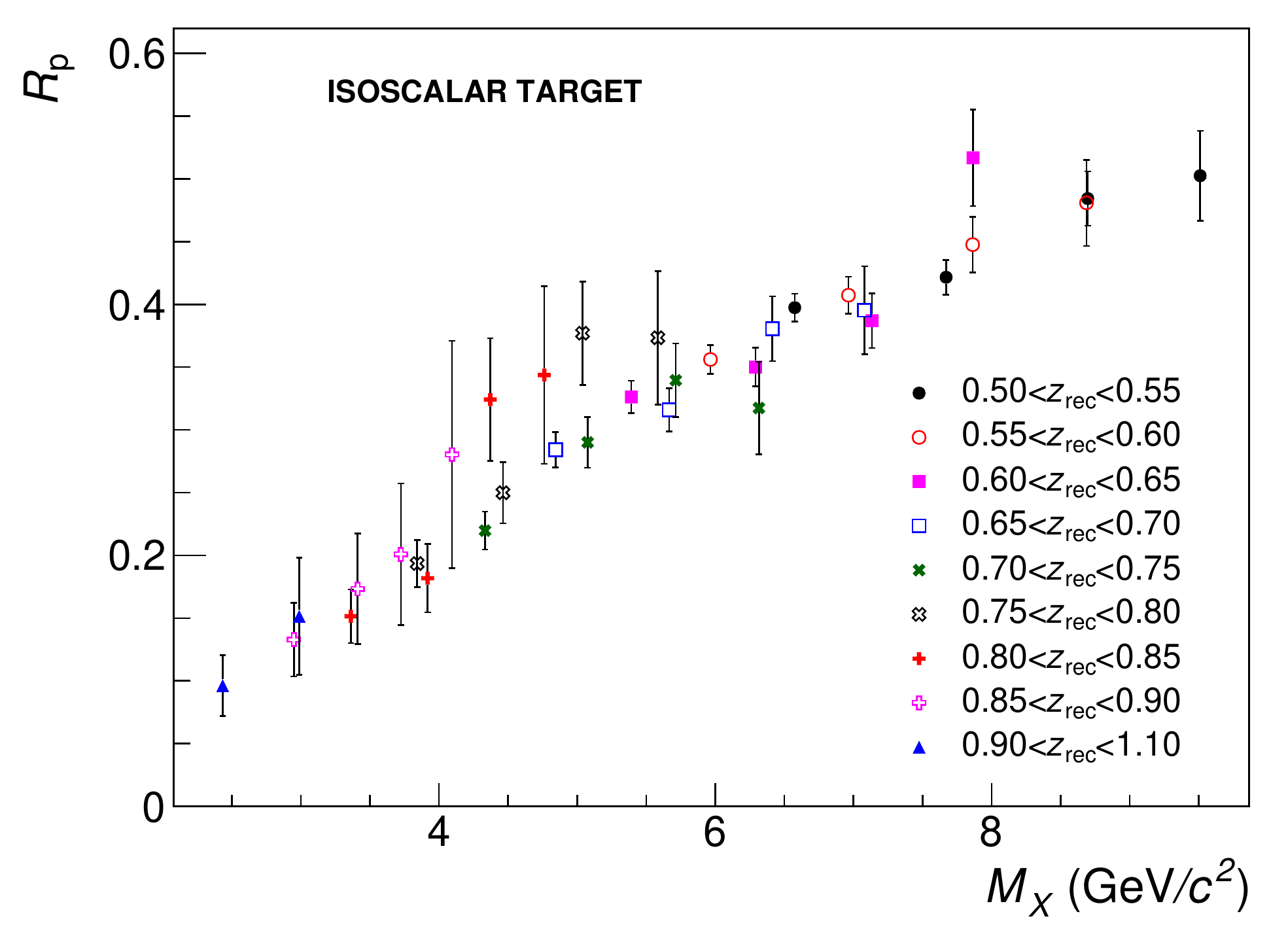}}
\caption{Results on $\rp$ as a function	of missing mass $M_{\rm X}$ for the first $x$-bin, $x<0.05$. For clarity only statistical uncertainties are shown. }
\label{fig:res3}
\end{figure}

In Ref.~\cite{comp_Kratio} it was found that the
$z$ and $\nu$  dependences, which are both 
unexpected in pQCD,
can be combined in the dependence on only 
one observable, which is the missing mass in the final state
that is approximately given by 
$M_{X}=\sqrt{M_{\rm p}^{2} + 2M_{\rm p} \nu(1-z) - Q^{2}(1-z)^2}$.
In Fig.~\ref{fig:res3}
the antiproton over proton multiplicity ratio $\rp$ 
is shown
as a function of the missing mass, and indeed
a smooth trend with overlapping points at different values of $z$ is observed.

\begin{figure}
\centerline{\includegraphics[clip,width=0.9\textwidth]{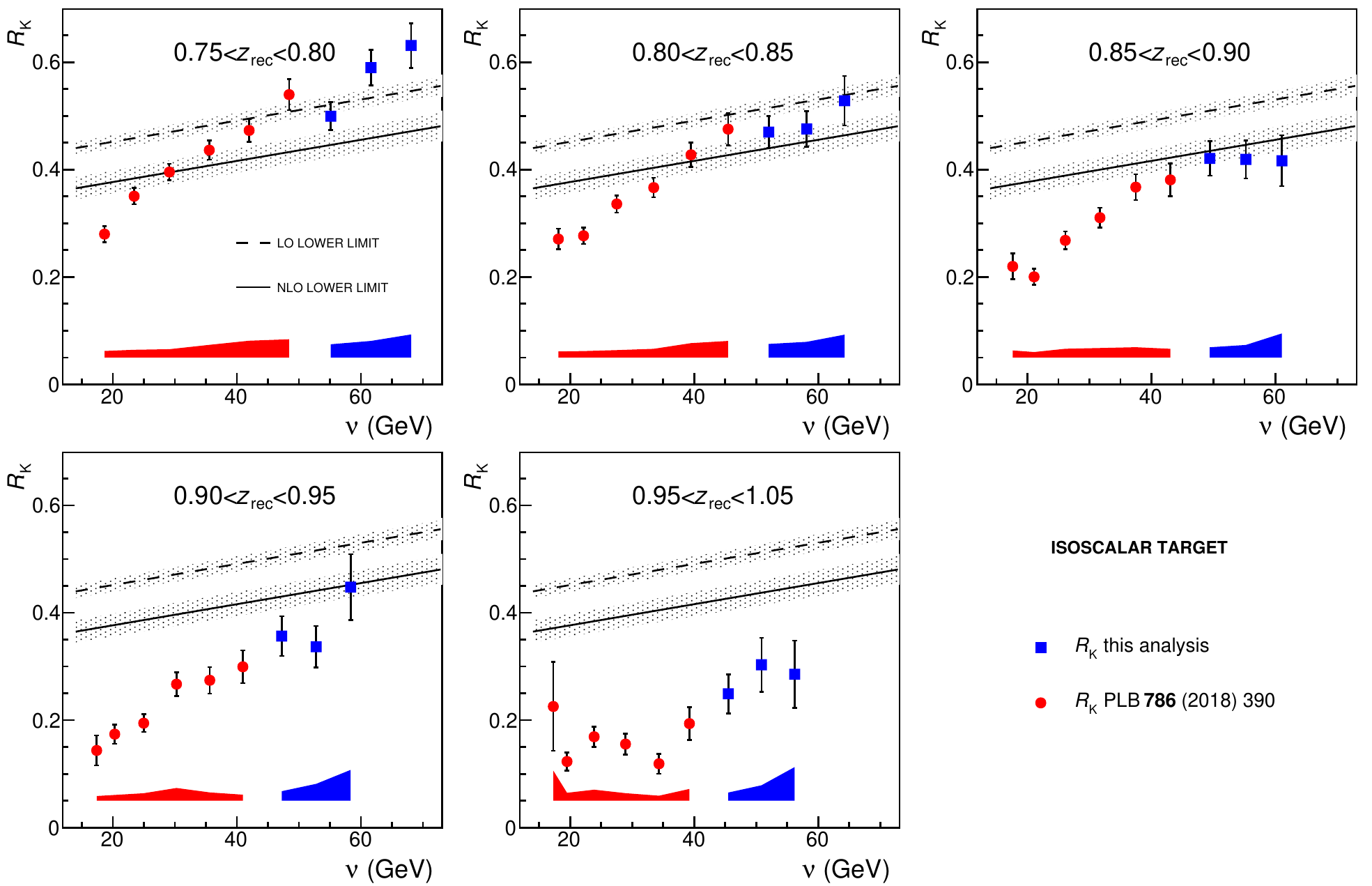}}
\caption{The K$^{-}$ over K$^+$ multiplicity ratio as a function of $\nu$ in five bins of $z$ obtained
in this analysis (blue) and in Ref.~\cite{comp_Kratio} (red). 
The errors bars represent statistical uncertainties.
The systematic uncertainties of the data points are indicated by the
shaded band at the bottom of each panel. 
The shaded bands around the (N)LO lower limits indicate their uncertainties.
}
\label{fig:res_rk1}
\end{figure}

The strong $\nu$ dependence of $\rk$ discussed above, 
as originally seen in Ref.~\cite{comp_Kratio},
was also the inspiration to extend the covered $\nu$
range by improving the RICH K-$\pi$ separation.  
In this way, kaon identification up to 55 GeV/$c$ 
was achieved  instead of 40 GeV/$c$ previously, which allows us to extend the covered $\nu$
range in every $z$ bin. In Fig.~\ref{fig:res_rk1}, the obtained results of $\rk$ in bins of $z$ 
as a function of $\nu$ in the extended momentum range are compared 
to the ones published in Ref.~\cite{comp_Kratio}, as well as to the NLO pQCD lower limit for $\rk$.
The results confirm that the compatibility with pQCD expectations is better at higher $\nu$. They also suggest that 
with increasing values of $\nu$ the growth of the ratio $\rk$ becomes smaller. These results are also given in Table~\ref{tab:res2}.

\begin{figure}
\centerline{\includegraphics[clip,width=0.9\textwidth]{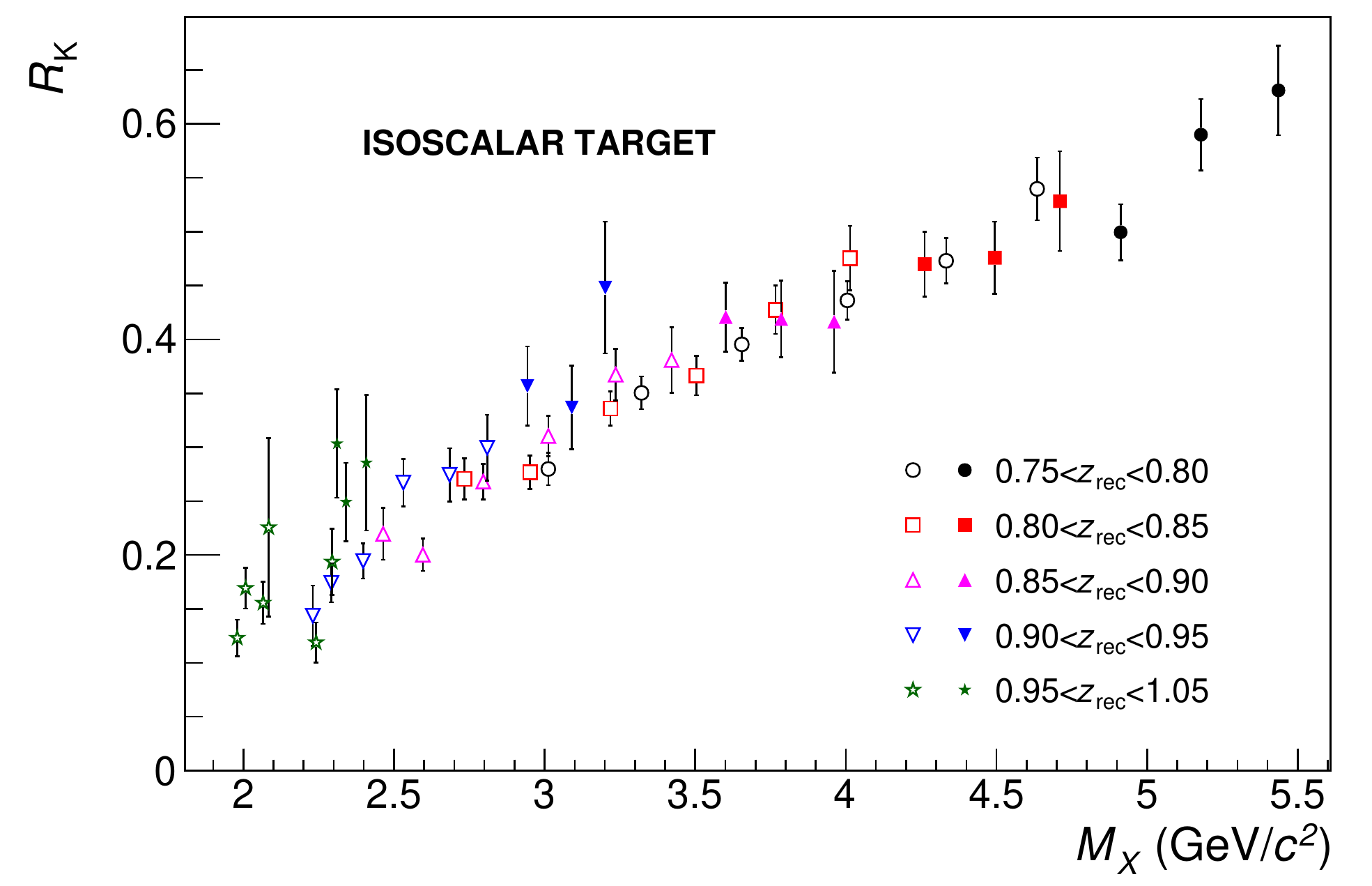}}
\caption{The K$^{-}$ over K$^+$ multiplicity ratio presented as a function of $M_X$
for this analysis (full symbols) and for the analysis in Ref.~\cite{comp_Kratio} (open symbols),
see text for details. For 
clarity only statistical uncertainties are shown.
}
\label{fig:res_rk2}
\end{figure}

For completeness, in Fig.~\ref{fig:res_rk2} the values of $\rk$ in the extended momentum range are compared to our earlier results \cite{comp_Kratio} as a function of missing mass.
The smooth growth with $M_X$  is still seen over the full kinematically accessible range. 
Now there is larger 
overlap in $M_X$ between different $z$-bins,
$i.e.$ one can find $M_X$ regions
where in four different $z$ bins at very different 
values of $\nu$ 
the results on $\rk$ are found to be consistent 
with one another.

\begin{table}[ht]
\begin{center}
\caption{Extracted values of $\rp$ with statistical and systematic uncertainties,
bin limits of $z$ $(z_{\rm min}, z_{\rm max})$, and 
average values of $x$, $Q^2$, $z_{\rm rec}$ and $z_{\rm corr}$ in the  first (upper part) and second
(lower part) $x$-bin.}
\begin{tabular}{ c|c|c|c|c|c|c|c }
bin &  $x $ & $ Q^2 $ (GeV/$c)^2$   & $z_{\rm min}$  & $z_{\rm max}$ & $ z_{\rm rec} $ &  $ z_{\rm corr} $   & 
$\rp \pm \delta R_{\rm p, \, stat.} \pm \delta R_{\rm p, \, syst.}$ \\ \hline
1 & 0.021 & 2.4 & 0.50 & 0.55 & 0.524 & 0.524 & $ 0.4238 \pm 0.0078 \pm 0.0270 $ \\
2 & 0.022 & 2.2 & 0.55 & 0.60 & 0.575 & 0.575 & $ 0.3953 \pm 0.0082 \pm 0.0244 $ \\
3 & 0.022 & 2.1 & 0.60 & 0.65 & 0.624 & 0.624 & $ 0.3601 \pm 0.0089 \pm 0.0224 $ \\
4 & 0.023 & 2.0 & 0.65 & 0.70 & 0.675 & 0.675 & $ 0.3216 \pm 0.0098 \pm 0.0205 $ \\
5 & 0.024 & 1.9 & 0.70 & 0.75 & 0.724 & 0.724 & $ 0.2729 \pm 0.0109 \pm 0.0178 $ \\
6 & 0.025 & 1.8 & 0.75 & 0.80 & 0.775 & 0.774 & $ 0.2636 \pm 0.0141 \pm 0.0187 $ \\
7 & 0.026 & 1.8 & 0.80 & 0.85 & 0.826 & 0.820 & $ 0.2117 \pm 0.0165 \pm 0.0166 $ \\
8 & 0.026 & 1.7 & 0.85 & 0.90 & 0.878 & 0.865 & $ 0.1720 \pm 0.0224 \pm 0.0123 $ \\
9 & 0.028 & 1.7 & 0.90 & 1.10 & 0.948 & 0.915 & $ 0.1130 \pm 0.0220 \pm 0.0068 $ \\ \hline
1'& 0.100 &10.5 & 0.50 & 0.55 & 0.525 & 0.525 & $ 0.2646 \pm 0.0117 \pm 0.0176 $ \\
2'& 0.101 & 9.7 & 0.55 & 0.60 & 0.575 & 0.575 & $ 0.2448 \pm 0.0116 \pm 0.0183 $ \\
3'& 0.101 & 9.0 & 0.60 & 0.65 & 0.625 & 0.625 & $ 0.2072 \pm 0.0111 \pm 0.0174 $ \\
4'& 0.101 & 8.4 & 0.65 & 0.70 & 0.675 & 0.675 & $ 0.1941 \pm 0.0122 \pm 0.0158 $ \\
5'& 0.100 & 7.8 & 0.70 & 0.75 & 0.725 & 0.725 & $ 0.1824 \pm 0.0140 \pm 0.0170 $ \\
6'& 0.102 & 7.5 & 0.75 & 0.80 & 0.774 & 0.771 & $ 0.1405 \pm 0.0148 \pm 0.0173 $ \\
7'& 0.102 & 7.1 & 0.80 & 0.85 & 0.823 & 0.815 & $ 0.1659 \pm 0.0233 \pm 0.0210 $ \\
8'& 0.099 & 6.4 & 0.85 & 0.90 & 0.872 & 0.855 & $ 0.0991 \pm 0.0241 \pm 0.0125 $ \\
9'& 0.104 & 5.9 & 0.90 & 1.10 & 0.948 & 0.910 & $ 0.0615 \pm 0.0218 \pm 0.0078 $ \\
\end{tabular}
\label{tab:res0}
\end{center}
\end{table}

\begin{table}[ht]
\begin{center}
\caption{Extracted values of $\rp$ with statistical and systematic uncertainties, bin range of proton momenta ($p_{\rm rg}$ (GeV/$c$)),
bin range in $z$ $(z_{\rm rg})$, and 
average values of $x$, $Q^2$, $z_{\rm rec}$ and $z_{\rm corr}$ in the first $x$-bin.}
\begin{tabular}{ c|c|c|c|c|c|c|c }
bin &  $x $ & $ Q^2 $ (GeV/$c)^2$   &  $p_{\rm rg}$ (GeV/$c$) & $z_{\rm rg}$ & $ z_{\rm rec} $ &  $ z_{\rm corr} $   & 
$\rp \pm \delta R_{\rm p, \, stat.} \pm \delta R_{\rm p, \, syst.}$ \\ \hline
1a	&	0.022	&	1.9	&	20--30	&	0.50--0.55	&	0.524	&	0.524	& $	0.3973	\pm	0.0110	\pm	0.0260	$ \\
1b	&	0.020	&	2.5	&	30--40	&	0.50--0.55	&	0.525	&	0.525	& $	0.4215	\pm	0.0138	\pm	0.0262	$ \\
1c	&	0.020	&	3.2	&	40--50	&	0.50--0.55	&	0.524	&	0.524	& $	0.4843	\pm	0.0217	\pm	0.0289	$ \\
1d	&	0.021	&	4.0	&	50--60	&	0.50--0.55	&	0.526	&	0.526	& $	0.5024	\pm	0.0358	\pm	0.0357	$ \\ \hline
2a	&	0.023	&	1.9	&	20--30	&	0.55--0.60	&	0.575	&	0.575	& $	0.3561	\pm	0.0114	\pm	0.0218	$ \\
2b	&	0.020	&	2.3	&	30--40	&	0.55--0.60	&	0.574	&	0.574	& $	0.4073	\pm	0.0147	\pm	0.0244	$ \\
2c	&	0.020	&	3.0	&	40--50	&	0.55--0.60	&	0.575	&	0.575	& $	0.4476	\pm	0.0221	\pm	0.0287	$ \\
2d	&	0.020	&	3.6	&	50--60	&	0.55--0.60	&	0.575	&	0.575	& $	0.4808	\pm	0.0343	\pm	0.0362	$ \\ \hline
3a	&	0.024	&	1.8	&	20--30	&	0.60--0.65	&	0.624	&	0.624	& $	0.3262	\pm	0.0129	\pm	0.0196	$ \\
3b	&	0.021	&	2.1	&	30--40	&	0.60--0.65	&	0.625	&	0.625	& $	0.3499	\pm	0.0154	\pm	0.0213	$ \\
3c	&	0.020	&	2.7	&	40--50	&	0.60--0.65	&	0.624	&	0.624	& $	0.3870	\pm	0.0218	\pm	0.0266	$ \\
3d	&	0.020	&	3.3	&	50--60	&	0.60--0.65	&	0.624	&	0.624	& $	0.5167	\pm	0.0384	\pm	0.0401	$ \\ \hline
4a	&	0.026	&	1.7	&	20--30	&	0.65--0.70	&	0.675	&	0.675	& $	0.2840	\pm	0.0141	\pm	0.0171	$ \\
4b	&	0.021	&	2.0	&	30--40	&	0.65--0.70	&	0.675	&	0.675	& $	0.3160	\pm	0.0173	\pm	0.0200	$ \\
4c	&	0.020	&	2.5	&	40--50	&	0.65--0.70	&	0.675	&	0.675	& $	0.3806	\pm	0.0258	\pm	0.0280	$ \\
4d	&	0.020	&	3.1	&	50--60	&	0.65--0.70	&	0.675	&	0.675	& $	0.3954	\pm	0.0350	\pm	0.0317	$ \\ \hline
5a	&	0.027	&	1.7	&	20--30	&	0.70--0.75	&	0.724	&	0.724	& $	0.2197	\pm	0.0151	\pm	0.0132	$ \\
5b	&	0.022	&	1.9	&	30--40	&	0.70--0.75	&	0.724	&	0.724	& $	0.2899	\pm	0.0202	\pm	0.0186	$ \\
5c	&	0.020	&	2.3	&	40--50	&	0.70--0.75	&	0.725	&	0.725	& $	0.3395	\pm	0.0293	\pm	0.0282	$ \\
5d	&	0.020	&	2.9	&	50--60	&	0.70--0.75	&	0.724	&	0.724	& $	0.3174	\pm	0.0368	\pm	0.0279	$ \\ \hline
6a	&	0.028	&	1.6	&	20--30	&	0.75--0.80	&	0.776	&	0.773	& $	0.1935	\pm	0.0188	\pm	0.0124	$ \\
6b	&	0.022	&	1.9	&	30--40	&	0.75--0.80	&	0.775	&	0.774	& $	0.2499	\pm	0.0243	\pm	0.0188	$ \\
6c	&	0.020	&	2.2	&	40--50	&	0.75--0.80	&	0.774	&	0.774	& $	0.3770	\pm	0.0411	\pm	0.0281	$ \\
6d	&	0.020	&	2.7	&	50--60	&	0.75--0.80	&	0.774	&	0.773	& $	0.3734	\pm	0.0532	\pm	0.0445	$ \\ \hline
7a	&	0.029	&	1.6	&	20--30	&	0.80--0.85	&	0.827	&	0.819	& $	0.1515	\pm	0.0214	\pm	0.0108	$ \\
7b	&	0.023	&	1.8	&	30--40	&	0.80--0.85	&	0.824	&	0.819	& $	0.1818	\pm	0.0272	\pm	0.0149	$ \\
7c	&	0.020	&	2.1	&	40--50	&	0.80--0.85	&	0.824	&	0.821	& $	0.3242	\pm	0.0489	\pm	0.0202	$ \\
7d	&	0.020	&	2.5	&	50--60	&	0.80--0.85	&	0.823	&	0.820	& $	0.3437	\pm	0.0707	\pm	0.0560	$ \\ \hline
8a	&	0.030	&	1.5	&	20--30	&	0.85--0.90	&	0.882	&	0.866	& $	0.1329	\pm	0.0293	\pm	0.0088	$ \\
8b	&	0.024	&	1.8	&	30--40	&	0.85--0.90	&	0.875	&	0.862	& $	0.1733	\pm	0.0440	\pm	0.0124	$ \\
8c	&	0.020	&	2.0	&	40--50	&	0.85--0.90	&	0.874	&	0.865	& $	0.2008	\pm	0.0564	\pm	0.0127	$ \\
8d	&	0.020	&	2.3	&	50--60	&	0.85--0.90	&	0.872	&	0.865	& $	0.2802	\pm	0.0906	\pm	0.0437	$ \\ \hline
9ab	&	0.031	&	1.5	&	20--40	&	0.90--1.10	&	0.954	&	0.917	& $	0.0958	\pm	0.0242	\pm	0.0057	$ \\
9cd	&	0.022	&	2.0	&	40--60	&	0.90--1.10	&	0.936	&	0.910	& $	0.1466	\pm	0.0451	\pm	0.0099	$ \\
\end{tabular}
\label{tab:res1}
\end{center}
\end{table}

\begin{table}[ht]
\begin{center}
\caption{Extracted values of $\rk$ with statistical and systematic uncertainties, bin range of kaon momenta ($p_{\rm rg}$ (GeV/$c$)),
bin range in $z$ $(z_{\rm rg})$, and 
average values of $x$, $Q^2$, $z_{\rm rec}$ and $z_{\rm corr}$.}
\begin{tabular}{c|c|c|c|c|c|c|c}
bin &  $x $ & $ Q^2 $ (GeV/$c)^2$   &  $p_{\rm rg}$ (GeV/$c$) & $z_{\rm rg}$ & $ z_{\rm rec} $ &  $ z_{\rm corr} $   &
$\rk \pm \delta R_{\rm K, \, stat.} \pm \delta R_{\rm K, \, syst.}$ \\ \hline
1g& 0.021&	2.1& 40--45 & 0.75--0.80&	0.774&	0.774& $0.4994 \pm 0.0260\pm 0.0246 $ \\
1h& 0.020&	2.3& 45--50 & 0.75--0.80&	0.774&	0.774& $0.5899 \pm 0.0332\pm 0.0308 $ \\
1i& 0.019&	2.4& 50--55 & 0.75--0.80&	0.774&	0.774& $0.6310 \pm 0.0415\pm 0.0429 $ \\ \hline
2g& 0.022&	2.1& 40--45 & 0.80--0.85&	0.824&	0.822& $0.4697 \pm 0.0300\pm 0.0252 $ \\
2h& 0.020&	2.2& 45--50 & 0.80--0.85&	0.823&	0.822& $0.4757 \pm 0.0334\pm 0.0291 $ \\
2i& 0.019&	2.3& 50--55 & 0.80--0.85&	0.823&	0.822& $0.5282 \pm 0.0462\pm 0.0425 $ \\ \hline
3g& 0.022&	2.0& 40--45 & 0.85--0.90&	0.872&	0.868& $0.4207 \pm 0.0320\pm 0.0190 $ \\
3h& 0.021&	2.1& 45--50 & 0.85--0.90&	0.873&	0.869& $0.4190 \pm 0.0356\pm 0.0233 $ \\
3i& 0.020&	2.2& 50--55 & 0.85--0.90&	0.872&	0.869& $0.4164 \pm 0.0473\pm 0.0448 $ \\ \hline
4g& 0.022&	1.9& 40--45 & 0.90--0.95&	0.921&	0.911& $0.3567 \pm 0.0368\pm 0.0178 $ \\
4h& 0.021&	2.1& 45--50 & 0.90--0.95&	0.921&	0.911& $0.3368 \pm 0.0388\pm 0.0315 $ \\
4i& 0.020&	2.2& 50--55 & 0.90--0.95&	0.921&	0.913& $0.4480 \pm 0.0611\pm 0.0575 $ \\ \hline
5g& 0.023&	1.9& 40--45 & 0.95--1.05&	0.974&	0.945& $0.2492 \pm 0.0363\pm 0.0153 $ \\
5h& 0.022&	2.0& 45--50 & 0.95--1.05&	0.975&	0.952& $0.3033 \pm 0.0502\pm 0.0288 $ \\
5i& 0.020&	2.1& 50--55 & 0.95--1.05&	0.974&	0.952& $0.2856 \pm 0.0628\pm 0.0628 $ \\ 
\end{tabular}
\label{tab:res2}
\end{center}
\end{table}

\section{Summary}

In this article the $\bar{\rm p}$ over p multiplicity
ratio $\rp$, obtained from semi-inclusive measurements of deep-inelastic lepton-nucleon scattering
at 
$z$-values above 0.5, 
is presented for the first time. 
In the whole studied $z$-region the 
ratio $\rp$ is observed to be below the lower limit predicted by LO pQCD. It is found to be significantly 
smaller than the K$^-$ over K$^+$ multiplicity ratio $\rk$
as presented in our previous letter, while
in naive LO pQCD both ratios are expected to be very  similar.
A strong dependence on the virtual-photon energy $\nu$ is observed,
which is also not expected
by LO pQCD but was already seen for the ratio $\rk$ 
in our earlier analysis.
In this article, the analysis of  $\rk$ is 
extended to  larger values of $\nu$ up to 70 GeV. 
The obtained results suggest that for high $\nu$ values there is an indication for  saturation of $\rk$
at or above the value predicted by NLO pQCD.
The present studies provide further support 
that the additional correction to the pQCD formalism 
suggested in our previous paper,
which takes into account the phase space available for hadronisation, 
depends on the mass of the produced hadron.

\section*{Acknowledgements}
We gratefully acknowledge the support of the CERN management and staff and the
skill and effort of the technicians of our collaborating institutes. 
This work was made possible by the financial support of our funding agencies.


\begin{thebibliography}{99}

\bibitem{dglap} V.N. Gribov and L.N. Lipatov, Sov. J. Nucl. Phys. {\bf 15} (1972) 438; L.N. Lipatov, {\it ibid.} {\bf 20} (1975) 95; G. Altarelli and G. Parisi, Nucl. Phys. B {\bf 126} (1977) 298; Yu.L. Dokshitzer, Sov. Phys. JETP {\bf 46} (1977) 641.
\bibitem{hkns07} M. Hirai, S. Kumano, T.-H. Nagai and K. Sudoh, Phys. Rev. D {\bf 75} (2007) 094009.
\bibitem{hkks16} M. Hirai, H. Kawamura, S. Kumano and K. Saito,
Prog. Theor. Exp. Phys. {\bf 2016} (2016) 113B04.
\bibitem{jam01}	N. Sato {\it et al.}, Phys. Rev. D {\bf 94} (2016) 114004.       
\bibitem{dss01} D. de Florian, R. Sassot and  M. Stratmann, Phys. Rev. D {\bf 75} (2007) 114010.
\bibitem{dss02} D. de Florian {\it et al.}, Phys. Rev. D {\bf 95} (2017) 094019.
\bibitem{nnpdf} NNPDF Collaboration, V. Bertone {\it et al.}, Eur. Phys. J. C  {\bf 77} (2017) 516.
\bibitem{hermes} HERMES Collaboration, A. Airapetian {\it et al.}, Phys. Rev. D {\bf 87} (2013) 074029.
\bibitem{comp_pi} COMPASS Collaboration, C. Adolph {\it et al.}, Phys. Lett. B {\bf 764} (2017) 1.
\bibitem{comp_K} COMPASS Collaboration, C. Adolph {\it et al.}, Phys. Lett. B {\bf 767} (2017) 133.
\bibitem{comp_hpt} COMPASS Collaboration, M. Aghasyan {\it et al.}, Phys. Rev. D {\bf 97} (2018) 032006.

\bibitem{comp_Kratio} COMPASS Collaboration, R. Akhunzyanov {\it et al.}, Phys. Lett. B {\bf 786} (2018) 390.
\bibitem{tmds}  M. Anselmino {\it et al.}, Phys. Rev. D {\bf 71} (2005) 074006.
\bibitem{nlo_form} W.~Furmanski and R.~Petronzio, Z. Phys. C {\bf 11} (1982) 293.
\bibitem{dsv} D.~de Florian, M. Stratmann and W. Vogelsang, Phys. Rev. D {\bf 57} (1998) 5811.
\bibitem{diquark} R. Jakob, P. J. Mulders and J. Rodrigues, Nucl. Phys. A {\bf 626} (1997) 937.



\bibitem{pdf_mstw08} A.~D. Martin, W.~J. Stirling, R.~S. Thorne and G. Watt, Eur. \ Phys. \ J. \ C {\bf 64} (2009) 653.
\bibitem{pdf_mmht14} L.~A. Harland-Lang, A.~D. Martin, P. Motylinski and R.~S. Thorne, Eur. Phys. J. C {\bf 75} (2015) 204.
\bibitem{pdf_nnpdf} NNPDF Collaboration, R. D. Ball {\it et al.}, J. High En. Phys. {\bf 04} (2015) 040.
\bibitem{lepto} G. Ingelman, A. Edin and J. Rathsman, Comput.\ Phys.\ Commun.\  {\bf 101} (1997) 108.
\bibitem{lund} T. Sj\"ostrand, LU-TP-95-20, CERN-TH-7112-93-REV, hep-ph/9508391.
\bibitem{Aram} A. Kotzinian, Eur. Phys. J. C {\bf 44} (2005) 211.
\bibitem{TMC_old} J.~V.~Guerrero {\it et al.}, J. High En. Phys. {\bf 09} (2015) 169.
\bibitem{TMC_ChL} E.~Christova and E.~Leader, Phys.\ Rev.\ D {\bf 94} (2016) 096001.
\bibitem{TMC_new} J.~V. Guerrero and A. Accardi, Phys.\ Rev. \ D {\bf 97} (2018) 114012.
\bibitem{WVog_res} D.~P.~Anderle, F.~Ringer and W.~Vogelsang, Phys.\ Rev.\ D {\bf 87} (2013) 034014.
\bibitem{Ref2_1} D.~P. Anderle, M. Stratmann and F. Ringer, Phys.\ Rev.\ D {\bf 92} (2015) 114017.
\bibitem{Ref2_2} D.~P. Anderle, T. Kaufmann, M. Stratmann and F. Ringer, Phys.\ Rev.\ D {\bf 95} (2017) 054003.
\bibitem{Ref2_3} M. Epele, C.~G. Canal and R. Sassot, Phys.\ Rev.\ D {\bf 94} (2016) 034037.
\bibitem{Ref2_4} M. Epele, C.~G. Canal and R. Sassot, Phys. Lett. B {\bf 790} (2019) 102.
\bibitem{Ref2_5} A. Accardi and A. Signori, Phys. Lett. B {\bf 798} (2019) 134993.
\bibitem{nim} COMPASS Collaboration, P. Abbon  {\it et al.}, Nucl. Instr. and Meth. A {\bf 577} (2007) 455.
\bibitem{rich_pid} P. Abbon {\it et al.}, Nucl. Instr. and Meth. A {\bf 631} (2011) 26.
\bibitem{djangoh} E. C. Aschenauer {\it et al.},  Phys. Rev. D {\bf 88} (2013) 114025; 
H. Spiesberger, HERACLES and DJANGOH : Event Generation of ep Interactions at HERA Including Radiative Processes (version 1.6).


\end{thebibliography}
\end{document}